\documentclass{jfm}

\usepackage{graphicx}
\usepackage{epstopdf, epsfig}
\usepackage{hyperref}
\hypersetup{
  unicode=true, 
  pdftitle={Span effect on the turbulence nature of flow past a circular cylinder},
  pdfauthor={Bernat Font},
  pdfnewwindow=true, 
  colorlinks=true,  
  pdfborder={0 0 0},  
  linkcolor=blue,
  linktoc=all,    
  citecolor=blue,
  urlcolor=blue,
  breaklinks=false,
}

\graphicspath{{./img/}}

\shorttitle{Span effect on the turbulence nature of flow past a circular cylinder}
\shortauthor{B. Font Garcia, G. D. Weymouth, V.-T. Nguyen and O. R. Tutty}

\title{Span effect on the turbulence nature of flow past a circular cylinder}

\author{Bernat Font Garcia\aff{1,2},
		Gabriel D. Weymouth\aff{1}
		\corresp{\email{g.d.weymouth@soton.ac.uk}},
		Vinh-Tan Nguyen\aff{2},
  		\and Owen R. Tutty\aff{1}}

\affiliation{\aff{1}Faculty of Engineering Physical Sciences, University of Southampton, SO17 1BJ Southampton, UK
\aff{2}Institute of High Performance Computing, Singapore Agency for Science, Technology and Research (A*STAR), 138632, Singapore}

\begin{document}

\maketitle

\begin{abstract}
Turbulent flow evolution and energy cascades are significantly different in two-dimensional (2D) and three-dimensional (3D) flows. Studies have investigated these differences in obstacle-free turbulent flows, but solid boundaries have an important impact on the cross-over between 3D to 2D turbulence dynamics. In this work, we investigate the span effect on the turbulence nature of flow past a circular cylinder at $\Rey=10000$. It is found that even for highly anisotropic geometries, 3D small-scale structures detach from the walls. Additionally, the natural large-scale rotation of the K\'{a}rm\'{a}n vortices rapidly two-dimensionalises those structures if the span is 50\% of the diameter or less. We show this is linked to the span being shorter than the Mode B instability wavelength. The conflicting 3D small-scale structures and 2D K\'{a}rm\'{a}n vortices result in 2D and 3D turbulence dynamics which can coexist at certain locations of the wake depending on the domain geometric anisotropy.
\end{abstract}

\section{Introduction}
Incompressible viscous flow past two-dimensional (2D) bluff bodies involves complex physics such as the well-known von-K\'{a}rm\'{a}n street phenomenon as well as three-dimensional (3D) wake dynamics as the Reynolds number ($\Rey$) is increased \citep{Roshko1954, Williamson1996b, Williamson1996a}. Due to the two-dimensionality of circular cylinders, some authors have used the 2D Navier-Stokes equations on multiple planes located along the span of the cylinder as a simplified model of the three-dimensionality without increased computational cost (\textit{a.k.a.} strip theory method). Such strip theory methods are used in offshore and civil engineering applications to model flow along slender structures where the computational cost of fully-resolved simulations is prohibitive, such as marine risers, tow and mooring cable systems, and tall pillars. However, the physics inherent in 2D simulations lead to poor 3D predictions \citep{Bao2016}, and a better understanding of the evolution from 3D to 2D turbulent wakes is needed to improve strip theory methods.

The fluid mechanics of turbulent flows behave quite differently when the spanwise spatial dimension is much more constrained than the others. Instead of having a direct cascade of the turbulence kinetic energy (TKE) from the integral scales down to the dissipative scales, there is a dual cascade of (direct) enstrophy and (inverse) energy. This was first suggested by \cite{Kraichnan1967} and further developed by \cite{Leith1968} and \cite{Batchelor1969} (known as the KLB 2D turbulence theory). More recently, the dual cascade has been demonstrated both experimentally and computationally \citep[for a comprehensive review see][]{Boffetta2012}. Studies such as \cite{Xiao2009} have shown that physical processes such as vortex-thinning and vortex-merging dominate the dynamics of 2D turbulence generating larger and more intense vortical structures the energy of which piles up at the integral scale for bounded domains. 

\begin{figure}
\vspace*{0.2cm}
  \centerline{\includegraphics[width=0.4\textwidth]{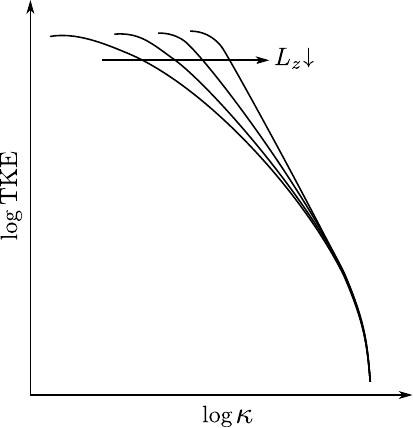}}
  \caption{Sketch of the TKE cascade distribution across the scales as the span is reduced. By constricting the domain, the integral scale is reduced and the energy-containing scales can feed less energy to the inertial and dissipative scales.}
\label{fig:ek_sketch}
\end{figure}

Previous work has studied the transition between 2D and 3D dynamics in obstacle-free turbulent flows in detail. The effect of length scale $L_z$ constriction on the TKE distribution across the scales (or wavenumbers $\kappa$) is sketched in figure \ref{fig:ek_sketch}. By constricting the domain, the size of the 3D energy-containing structures (integral scale structures) is reduced. Since the energy-containing structures feed the inertial subrange structures down to the 3D small-scale dissipative structures, smaller integral scale structures result in less energy fed into the inertial subrange structures and, consequently, to the dissipative structures. On the 2D limit, no 3D dissipative structures are present and, because of the lack of a dissipation mechanism, the turbulent vortical structures can only merge. This creates larger structures promoting an inverse energy cascade as shown in \cite{Kraichnan1967, Leith1968, Batchelor1969}.

\cite{Smith1996} reviewed the aspect ratio depth effect together with a rotation effect of forced turbulence on a $L_x \times L_y \times L_z$ periodic box. It was found that the turbulence dimensionality of the flow depended not only on the geometry constriction ($A\equiv L_z/L_x$), but also on the rotation intensity $\Omega$. A critical ratio between the span and the turbulence forcing scale was revealed below which two-dimensionalisation occurred for non-rotating cases. Furthermore, it was found that higher rotation rates induced a more significant 2D turbulence behaviour and that direct and inverse energy cascades for small and large scales can coexist respectively.  \cite{Celani2010} found a similar splitting of the turbulence cascade for a critical value of the relative forcing on a depth-restricted periodic box.

These differences between 2D and 3D turbulence dynamics have a significant practical impact. For example, the forces induced on a cylinder are larger in magnitude and variability in 2D systems \citep{Mittal1995, Norberg2003}. In an attempt to dissipate the energised vortical structures and to prevent the vortex-merging dynamics of 2D turbulence, models based on the turbulent-viscosity hypothesis (\textit{a.k.a.} the Boussinesq hypothesis) are incorporated into 2D strip theory methods. However, these models assume that the anisotropic Reynolds stress tensor is proportional to the mean rate-of-strain tensor by the scalar turbulent viscosity, which has been proven inaccurate even for simple shear flows \citep[p. 94]{Pope2000}. A further complication for strip theory is that the presence of walls adds a production mechanism of 3D turbulence which is able to generate very fine 3D vortical structures even in highly anisotropic geometries.

A ``thick'' strip theory method proposed by \cite{Bao2016} showed how strips with a certain thickness are able develop to 3D turbulence when its span is larger than the wavelength of the Mode B instability of circular cylinders (about one diameter). In fact, this instability creates rib-like streamwise vortical structures along the main K\'{a}rm\'{a}n 2D vortices \citep{Noack1999}. Therefore, it can be argued that the two-dimensionalisation of the wake arises from the geometry constriction which prevents the rib-like vortices to develop when there is not room enough for its natural wavelength. However, the connection between wake and wall turbulence and the persistence of 3D turbulent structures in constricted span flows has not been fully explored.

This work studies the geometry constriction effect on the turbulence nature of a flow past a circular cylinder at $\Rey=10000$. To do this, a series of simulations ranging from $L_z=10$ to pure 2D planes have been considered. As discussed above, the inclusion of a body boundary provides an important change to the turbulence production mechanisms compared to previous research and novel information on the transition and cross-over between 3D and 2D turbulence for very constricted domains in wall-generated turbulent shear flows. Multiple turbulence statistics are presented for the wide range constricted wakes, providing new data on the transition from 3D to 2D turbulence.

With this intention, we have structured the current paper as follows: \S\ref{Solver} describes the governing equations and the numerical methods as well as the computational details of the simulations. In \S\ref{Results and discussion}, the turbulence nature of the wake for the different cases is analysed (similarly to \cite{Biancofiore2014}) and discussed with results such as wake visualisation, velocity temporal spectras at different locations, Lumley's triangle, TKE spatial plots and vortex-stretching analysis.

\section{Problem formulation}\label{Solver}

This study considers the flow past a circular cylinder with diameter $D$ aligned on the $z$ direction in a 3D $\left(35\times20\times L_z\right)D$ rectangular domain, where $L_z$ is the non-dimensional span. To study the span effect on the turbulence nature of the wake, the following cases have been considered: $L_z=10, \pi, 1, 0.5, 0.25, 0.1$ as well as a fully 2D case. 

\begin{figure}
	\vspace*{0.2cm}
  \centerline{\includegraphics[width=0.8\textwidth]{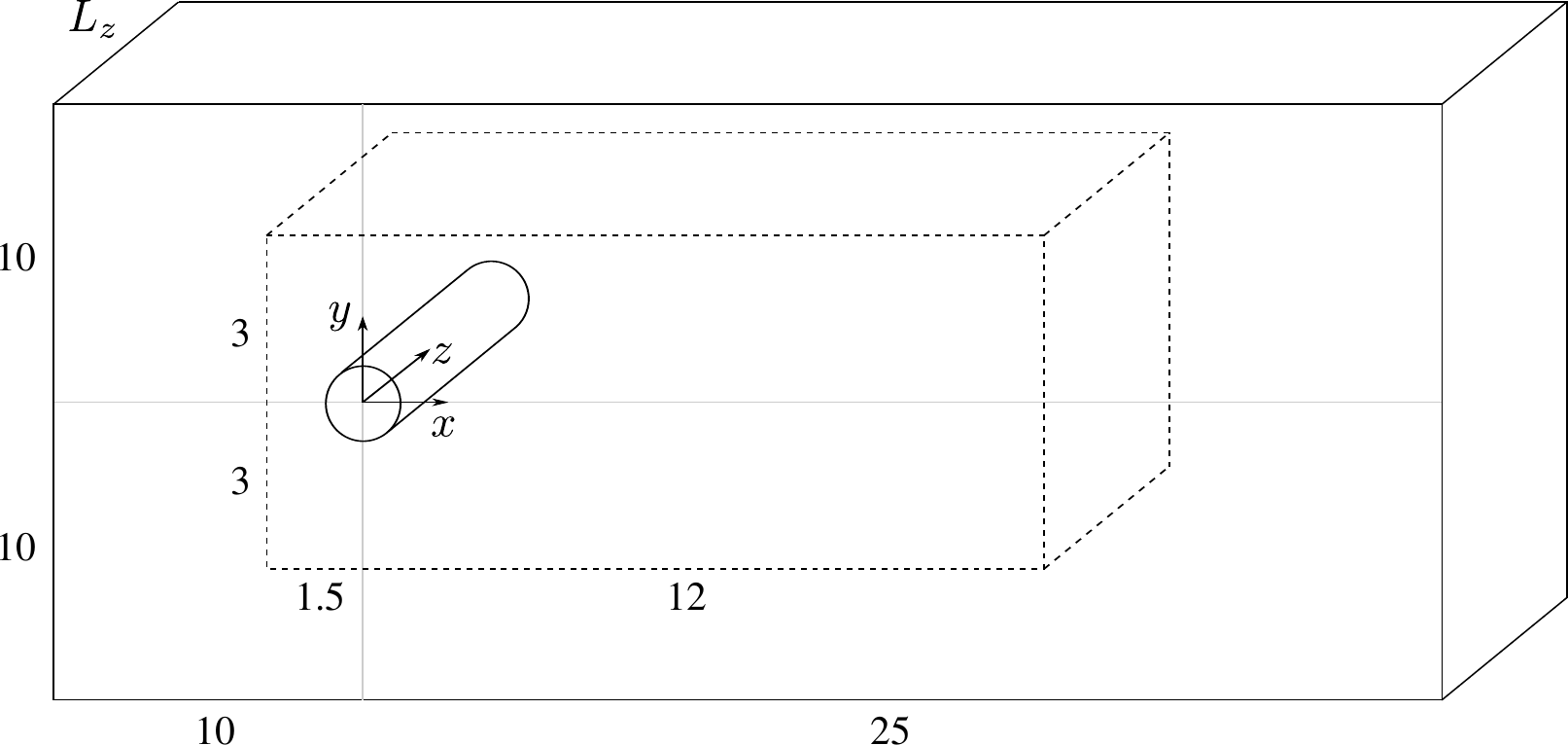}}
  \caption{Computational domain sketch non-dimensionalised with the cylinder diameter $D$. A very fine Cartesian grid domain (depicted in discontinuous lines) surrounds the cylinder and the close and mid wake regions. A stretched rectilinear grid (depicted in solid lines) transitions from the Cartesian grid to the boundaries.}
\label{fig:computational_domain}
\end{figure}

A Reynolds number of $\Rey=10000$ is selected. We define the Reynolds number as,
\begin{equation}
\Rey=\frac{UD}{\nu},
\end{equation}
where $U$ is the scaling velocity and $\nu$ is the kinematic viscosity of the fluid. Using this Reynolds number we ensure that the flow is well within the turbulent regime encompassing very different spatial and time scales.

\subsection{Governing equations and numerical method}

The incompressible viscous fluid motion is described by the continuity equation and the non-dimensional Navier-Stokes momentum equations,
\begin{eqnarray}
&\nabla\cdot\mathbf{u}=0,\label{eq:mass}\\
&\partial_{t}\mathbf{u}+\mathbf{u}\cdot\nabla \mathbf{u}=-\nabla p+\Rey^{-1}\nabla^2\mathbf{u},\label{eq:N-S}
\end{eqnarray}
where $\mathbf{u}\left(\mathbf{x}, t\right) = \left(u, v, w\right)$ is the velocity vector field, $p\left(\mathbf{x}, t\right)$ is the pressure field, $t$ is the time and $\mathbf{x}=\left(x,y,z\right)$ is the spatial vector. The initial condition is defined as $\mathbf{u}\left(\mathbf{x}, 0\right)=\left(1,0,0\right)$ in the fluid. The boundary conditions are: a uniform velocity profile on the inlet boundary, a natural convection condition on the outlet boundary, a no-penetration slip condition on the upper and lower boundaries, a periodic condition on the spanwise direction boundaries and a no-slip velocity condition on the cylinder. Periodic boundary conditions on the constricting planes are used as in previous studies on the cross-over of 2D and 3D turbulence for obstacle-free flows such as \cite{Smith1996}, \cite{Celani2010} and \cite{Biancofiore2014}. This choice is made to avoid the artificial high intensity turbulence enhancements and the deterioration of the 2D behaviour of the flow when a no-penetration condition is enforced on the constricting planes, as noted on \cite{Biancofiore2012} for obstacle-free turbulent wakes. Other studies such as \cite{Bao2016} and \cite{Bao2019} also use a periodic spanwise condition for thick strips modelling long circular cylinders.

The governing equations are numerically solved in a discrete rectilinear grid (details in \S\ref{subsec:comp_details}). We couple the discrete velocity and pressure fields with the Pressure-Poisson equation. The Boundary Data Immersion Method (BDIM) from \cite{Weymouth2011} has been used in our in-house implicit Large Eddy Simulation (ILES) solver. In short, the BDIM is an immersed boundary (IB) method which maps the governing equations of both the fluid motion and the solid motion into an interface providing an smooth transition of the system variables between the mediums. In this way, the system of fluid and body equations can be solved with a single meta-equation discretised on a rectilinear staggered grid. The BDIM has been extensively validated including cases of flow around bluff bodies such as \cite{Schulmeister2017} and \cite{Maertens2015}, where the latter also provides a detailed explanation of the numerical method. Our in-house code is second-order accurate in space (using the Quadratic Upstream Interpolation for Convective Kinematics, \textit{a.k.a.} QUICK, scheme) and second-order accurate in time (using a predictor-corrector algorithm). The implicit turbulence modelling derives from a flux-limited QUICK treatment of the convective terms, equivalent to optimum finite-volume schemes (for a review on ILES see \cite{Adams2009}). \cite{Hendrickson2019} have validated this ILES approach for intermediate Reynolds numbers similar to the $Re=10000$ used in the current work. 

\subsection{Computational details} \label{subsec:comp_details}
The domain is composed of a sufficiently fine Cartesian grid for the close and mid wake regions defined as  $\left(L_x\times L_y\times L_z\right)D$, where $L_x$ and $L_y$ are the non-dimensional horizontal and vertical lengths respectively. A stretched grid is considered for the regions far from the cylinder (see figure \ref{fig:computational_domain}). A resolution of 90 cells per diameter in all the spatial directions is chosen for the Cartesian grid subdomain. The resolution in all spatial directions is kept constant as the span is reduced.

The 3D simulations are started from a three-dimensionalised  2D flow snapshot. A time length of 200 units $\left(T = tU/D\right)$ is simulated before starting to record the flow statistics in order to achieve a statistically stationary state of the wake. The flow statistics are then recorded for a total of $500T$ (around 100 wake cycles). A verification and validation of the wake turbulence dynamics of the investigated test case is included in appendix \ref{appA}. Finally, the turbulence statistics of the $L_z=10$ and $L_z=\pi$ cases are very similar as displayed in figure \ref{fig:TKE}a. Hence, only the $L_z=\pi$ results are displayed on the other figures for clarity.

\section{Results and discussion}\label{Results and discussion}

\begin{figure}
	\vspace*{0.3cm}
  \centerline{\includegraphics[width=1\textwidth]{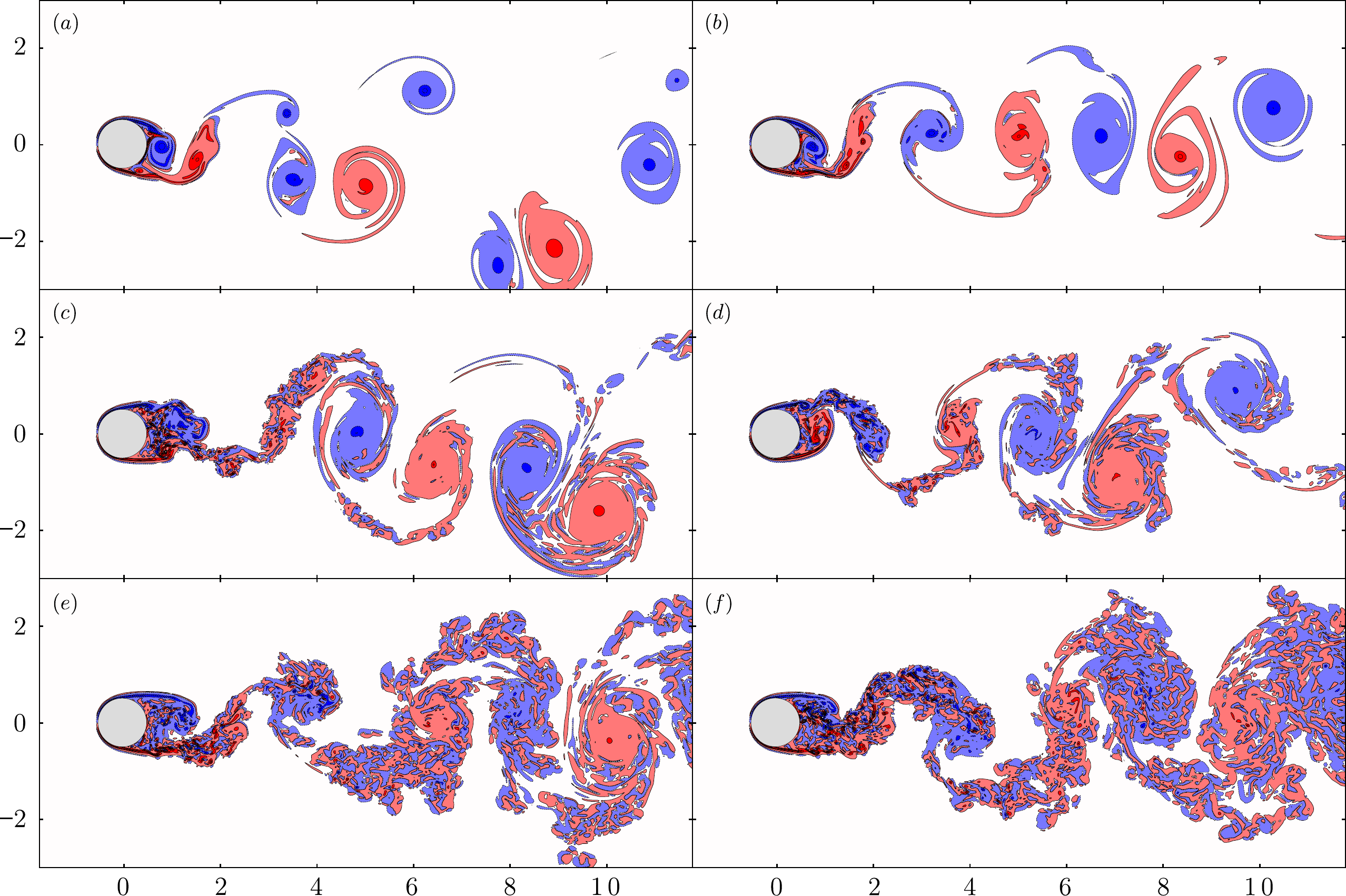}}
  \caption{Instantaneous vorticity $\omega_z$ (red is positive, blue is negative) at the $z=L_z/2$ plane for: $(a)$ $L_z=0$, $(b)$ $L_z=0.1$, $(c)$ $L_z=0.25$, $(d)$ $L_z=0.5$, $(e)$ $L_z=1$, $(f)$ $L_z=\pi$. }
\label{fig:instantaneous_vorticity}
\end{figure}

The flow field is displayed in figure \ref{fig:instantaneous_vorticity} in terms of the instantaneous vorticity component $\omega_z$ as the span is varied. The most striking feature is how the coherence of the K\'{a}rm\'{a}n vortices increases as the span is reduced. However, even in highly-anisotropic geometries such as $L_z=0.25$, small-scale 3D structures are generated from the cylinder wall. 

An important result is that the two-dimensionalisation of these structures is faster (in the sense that it occurs closer to the cylinder) as the domain is constricted because of the geometry constriction and the natural rotation of the K\'{a}rm\'{a}n vortices. The combination of these two mechanisms as a two-dimensionalisation method is also found in \cite{Smith1996} and \cite{Xia2011}. For the $L_z\geq1$ cases, the 3D small-scales structures detaching from the cylinder wall are not two-dimensionalised as rapidly. In fact, it can be appreciated that only the far wake region of the $L_z=1$ case displays a coherent K\'{a}rm\'{a}n vortex. This means that less anisotropic geometries promote a direct TKE cascade on the wake so that the 3D dissipative structures are still sustained far from the cylinder.

\begin{figure}
	\vspace*{0.3cm}
  \centerline{\includegraphics[width=1\textwidth]{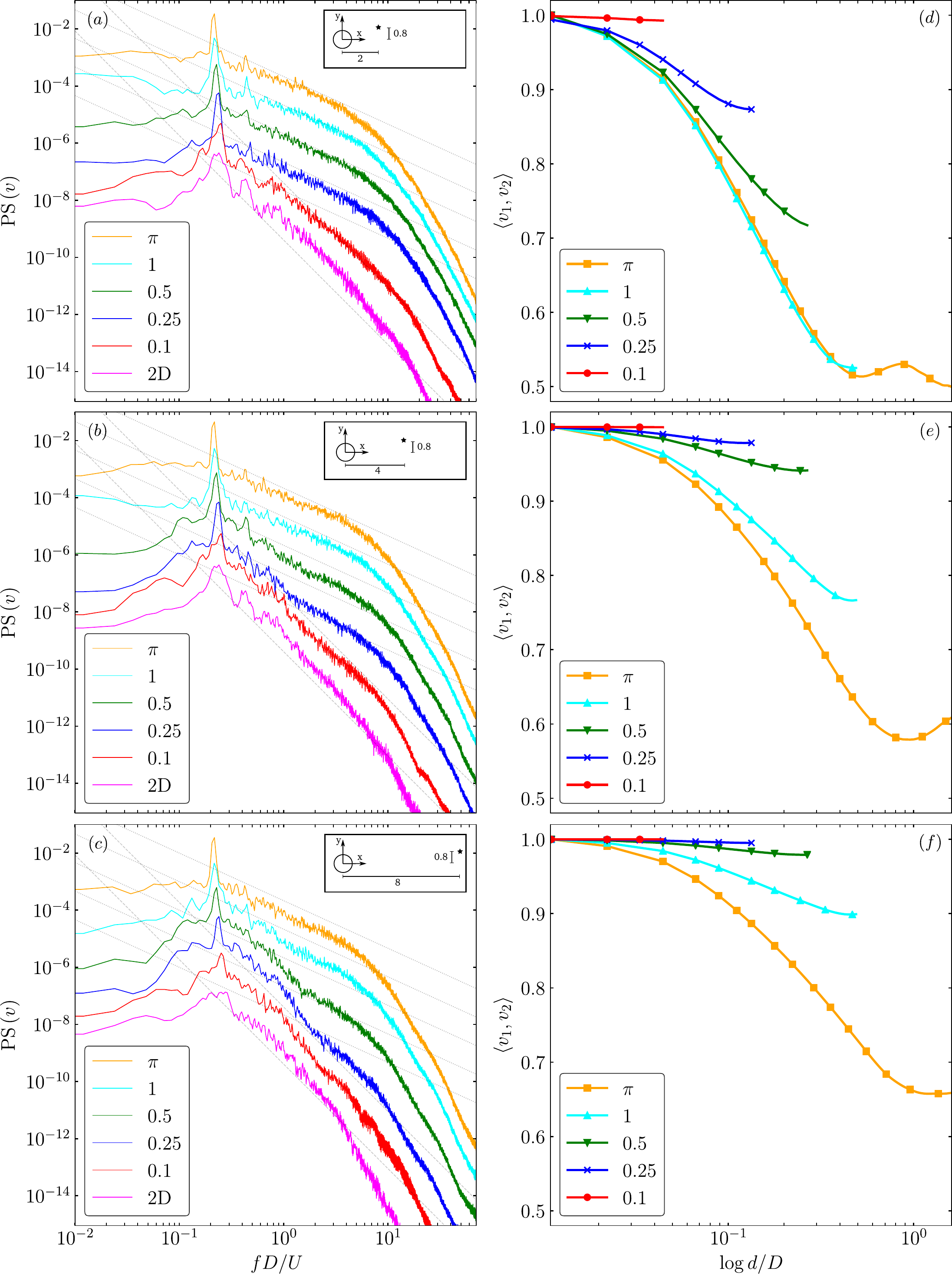}}
  \caption{Left: vertical velocity component temporal power spectra (PS) at different $\left((x,y\right)$ locations on the wake: $\left(a\right)=\left(2,0.8\right)$, $\left(b\right)=\left(4,0.8\right)$, $\left(c\right)=\left(8,0.8\right)$. The PS lines of each case are shifted a factor of 10 for clarity and the vertical axis ticks correspond to the $L_z=\pi$ case. The dashed lines have a $-11/3$ slope and the dotted lines have a $-5/3$ slope. Right: Two-point correlations along $z$ at the same $(x,y)$ locations as the left figures respectively. The correlation value for a given $d$ corresponds to the averaged value of the multiple correlations of pairs of points separated a distance $d$ along the span.}
\label{fig:velocity_spectras}
\end{figure}

Whether the wake turbulence dynamics are 2D or 3D is better captured on the TKE spectra which can be directly compared to classic turbulence theory. For this, the Taylor's hypothesis is considered and the temporal spectra at different points of the wake is calculated. Figure \ref{fig:velocity_spectras} a,b,c shows the temporal power spectra (PS) of the $v$ velocity component at $(x,y)=(2,0.8),(4,0.8),(8,0.8)$. The Welch method (using a time signal of 500 units split in 6 parts with 75\% overlapping) has been employed to compute the spectra at 8 different points along the span for each $(x,y)$ point. These spectras are then averaged resulting in a single spectra for each case.

First, note that all of the spectras display a peak around the $0.2$ non-dimensional frequency corresponding to the non-dimensional Strouhal number, $S_t=f_s D/U$, where $f_s$ is the vortex-shedding frequency. A smaller harmonic peak around $0.4$ non-dimensional frequency is also found for the spectra at $(x,y)=(2,0.8),(4,0.8)$. Second, the spectra at the closest analysed point to the cylinder (figure \ref{fig:velocity_spectras}a) displays a 3D turbulence behaviour with a $-5/3$ decaying rate with the exception of the pure 2D and the $L_z=0.1$ cases. For the latter cases, a decaying slope around $-11/3$ is found. These 2D-flow spectras are steeper than the $-3$ rate predicted by the classical 2D turbulence theory \citep{Kraichnan1967} because of the interaction of the coherent large-scale K\'{a}rm\'{a}n vortices, in agreement with the finds of \cite{Dritschel2008} and \cite{Biancofiore2014}. The filamentary vorticity (filaments of vorticity around the coherent vortices) is likely to be destroyed by the interaction of the large-scale vortices rather than viscous effect (specially for high $\Rey$ flows), thus limiting its range of scales. The coherent vortices induce a spiralling effect which limits the range of scales of incoherent filamentary vorticity \citep{Gilbert1988}. On the other hand, as a $-5/3$ decaying rate is captured for the $L_z=0.25$ case, it can be argued that 3D turbulence is being generated from the cylinder wall even for highly-reduced spans.

The $L_z=0.25$ and $L_z=0.5$ cases feature a decay rate that transitions from $-5/3$ to $-11/3$ as the spectras are computed further downstream from the cylinder. In particular, both cases have $-5/3$ slopes in \ref{fig:velocity_spectras}a and $-11/3$ in figure \ref{fig:velocity_spectras}c, quantifying the turbulent two-dimensionalisation in the rotating wake. Also note the coexistence of both 2D and 3D turbulence features for the $L_z=0.25,0.5$ and $L_z=1$ cases on figures \ref{fig:velocity_spectras}b and \ref{fig:velocity_spectras}c respectively. The low-frequency structures behave mostly 2D (decaying rate between $-3$ and $-11/3$, resulting from the presence of less coherent 2D structures than the pure 2D case) up to a certain point where a $-5/3$ rate is briefly recovered. Hence, high-frequency structures interact in a 3D fashion while low-frequency structures interact two-dimensionally. This finding is in agreement with \cite{Smith1996} and \cite{Celani2010}. 

Additionally, two-point correlations along the span have been analysed at the same $(x,y)$ locations as the PS plots (figure \ref{fig:velocity_spectras} d,e,f). Given a distance $d$ along $z$ ranging from 0 to $L_z/2$, the two-point correlation is calculated with the temporal signals of the vertical velocity component at multiple pairs of points (namely $v_1$ and $v_2$) as follows,
\begin{equation}
\left\langle v_1(\mathbf{x},t), v_2(\mathbf{x}+\mathbf{r},t) \right\rangle = \frac{\mathrm{cov}(v_1, v_2)}{\sqrt{\mathrm{cov}(v_1, v_1)\mathrm{cov}(v_2, v_2)}},
\end{equation}
where the distance vector is defined as $\mathbf{r}=(0,0,d)$. The multiple correlation coefficients for a given $d$ are then averaged corresponding to a data point in the plots. 

Very close to the cylinder (figure \ref{fig:velocity_spectras}d), the correlation coefficient quickly decreases with increasing $d$ for $L_z>0.1$. This indicates the presence of 3D structures near the body as also noted on the velocity spectra plot counterpart. Also, the decrease is more pronounced as the span increases. For $L_z=\pi$, it is worth noting a local correlation coefficient maximum around $d \approx 0.9$. This distance approximately corresponds to the Mode B instability wavelength $(\lambda_z)$. Since the rib-like vortices associated with Mode B instability (streamwise and cross-flow vorticity) are not very well defined at this $\Rey$ regime \citep{Chyu1996}, the correlation increase is not as significant as at lower $\Rey$. Still, $L_z=\pi$ is the only case displaying such phenomena because of the spanwise boundary conditions periodicity, which only allow the instability to develop if $L_z>\lambda_z$ (the $L_z=1$ case might be too critical to display such phenomenon considering also its intermittent nature). The correlation coefficient increases when calculated further downstream as shown in figure \ref{fig:velocity_spectras}e and \ref{fig:velocity_spectras}f, evidencing again the wake two-dimensionalisation.

\begin{figure}
	\vspace*{0.3cm}
  \centerline{\includegraphics[width=1\textwidth]{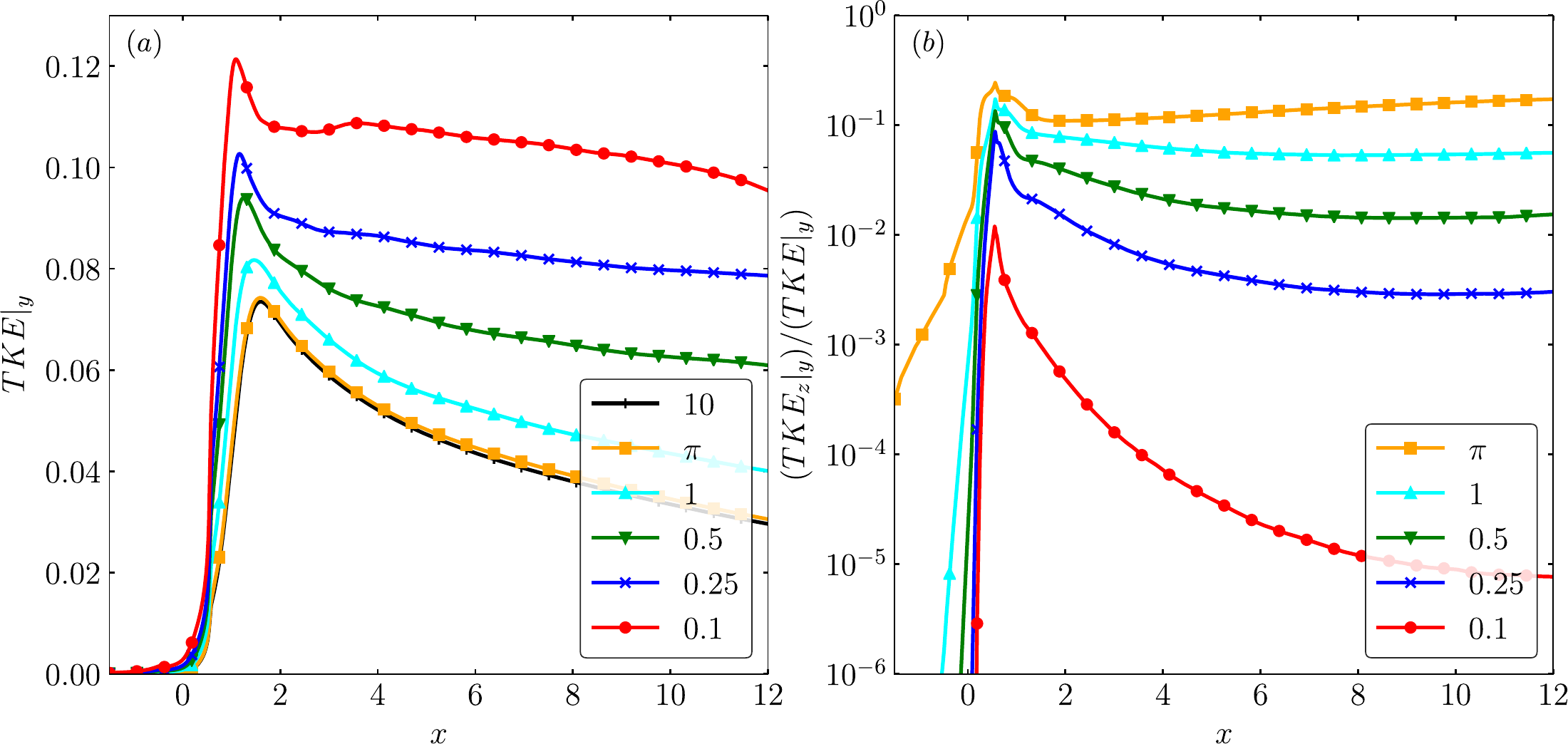}}
  \caption{TKE spatial plots. The TKE is computed from the normal Reynolds stresses and averaged on the vertical direction: $(a)$ total TKE, $(b)$ ratio of the spanwise component $\overline{w'w'}$ over the total TKE.}
\label{fig:TKE}
\end{figure}

In summary, the transition to a 2D wake is found at a certain point along the wake on all the cases with $L_z\le1$. The combination of the large-scale rotation from the K\'{a}rm\'{a}n vortices plus the geometry constriction are mainly responsible for this phenomenon. The main physical mechanism differing among the compared cases is the ability of the flow to develop Mode B-like 3D structures in the wake as a result of a sufficiently long span. The current $\Rey$ regime is characterised by a transition to turbulent flow at the shear layer (\textit{i.e.} the TrSL2 regime) as reported in \cite{Bloor1964} and \cite{Kourta1987}.  We argue that when the span is too short for the Mode B instability to develop and thereby sustain these 3D turbulent structures, the stratification effect of the K\'{a}rm\'{a}n vortices leads to the more coherent and energized wake seen in figure  \ref{fig:velocity_spectras} d,e,f.

Next, the TKE along the $x$ direction, the Lumley's triangle of turbulence and the ratio between the vortex-stretching and advection terms are examined to further support the observed phenomena. The TKE is defined as,
\begin{equation}
TKE = \frac{1}{2}\left(\overline{u'u'}+\overline{v'v'}+\overline{w'w'}\right),
\end{equation}
where $\overline{\cdot}$ denotes a time average plus a spanwise average and the subscript $\cdot'$ denotes a fluctuating quantity such as $a'=a-\overline{a}$. The six components of the Reynolds stress tensor $\overline{u_i' u_j'}$ have been computed using the following relation,
\begin{equation}
\overline{a'b'} = \overline{ab} - \overline{a}\overline{b}.
\end{equation}

Figure \ref{fig:TKE} shows the streamwise spatial distribution of the TKE averaged along the $y$ direction from $-L_y/2$ to $L_y/2$ (noted as $\cdot|_y$). From a general point of view, it can be observed that the total TKE (figure  \ref{fig:TKE}a) peaks right after the recirculation region noting that the latter increases slightly with the span. Also, the total TKE increases as the span is reduced because of the 2D vortex-merging processes that generate larger and more energised vortical structures. The contribution of the spanwise normal stress to the total TKE increases with the span as shown in figure \ref{fig:TKE}b. Also, a decay of $\overline{w'w'}$ right after being generated from the cylinder wall can be noted and it becomes faster as the span is constricted. 

\begin{figure}
 \vspace*{0.3cm}
  \centerline{\includegraphics[width=0.5\textwidth]{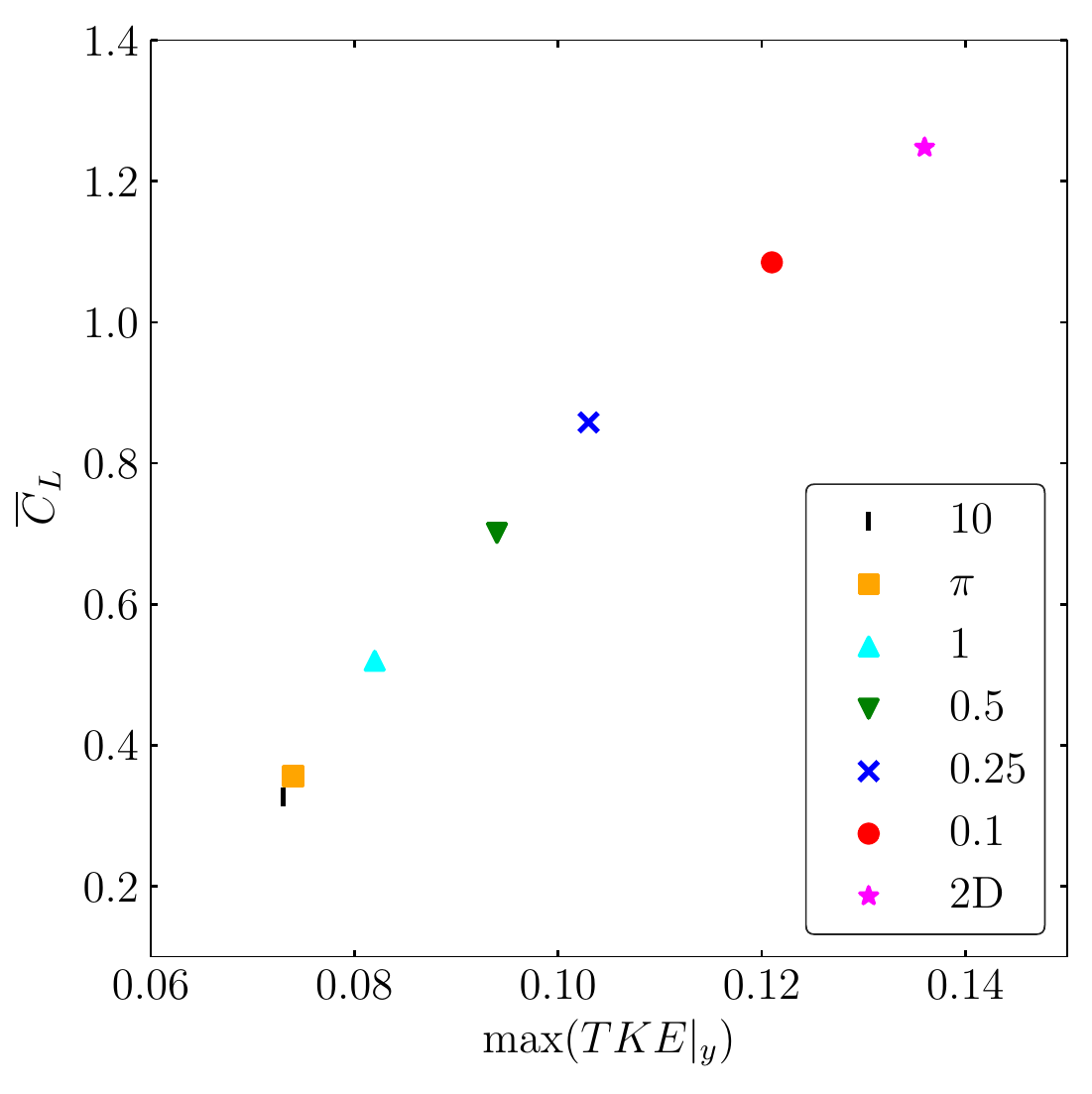}}
  \caption{Effect of the span constriction on the TKE and the $\overline{C}_L$. The latter is calculated as $\overline{C}_L=|2F_y/(\rho U^2 D L_z)|_{\mathrm{rms}}$, where $F_y$ is the vertical lift force and $\rho$ is the constant fluid density.}
\label{fig:TKE_CL}
\end{figure}

The effect of the span on the TKE compared to the lift coefficient r.m.s. value $\left(\overline{C}_L\right)$ displays a quasi-linear relation as shown in figure \ref{fig:TKE_CL}. Note that the drop on both values is more sensitive to the span constriction at the range where both 2D and 3D turbulence dynamics co-exist, \textit{i.e.} $0.25\le L_z\le1$. On the other hand, a very small change of both values can be appreciated when 3D turbulence fully dominates the wake, \textit{i.e.} $L_z\ge\pi$. Furthermore, as shown in figure \ref{fig:TKE}a, highly constricted domains yield large values of TKE because of the energised 2D large-scale vortical structures present at the wake. Even when small-scale 3D structures are present in the close wake region and coexist with the large 2D structures, the values of both the TKE and the $\overline{C}_L$ spike.

\begin{figure}
	\vspace*{0.3cm}
  \centerline{\includegraphics[width=0.8\textwidth]{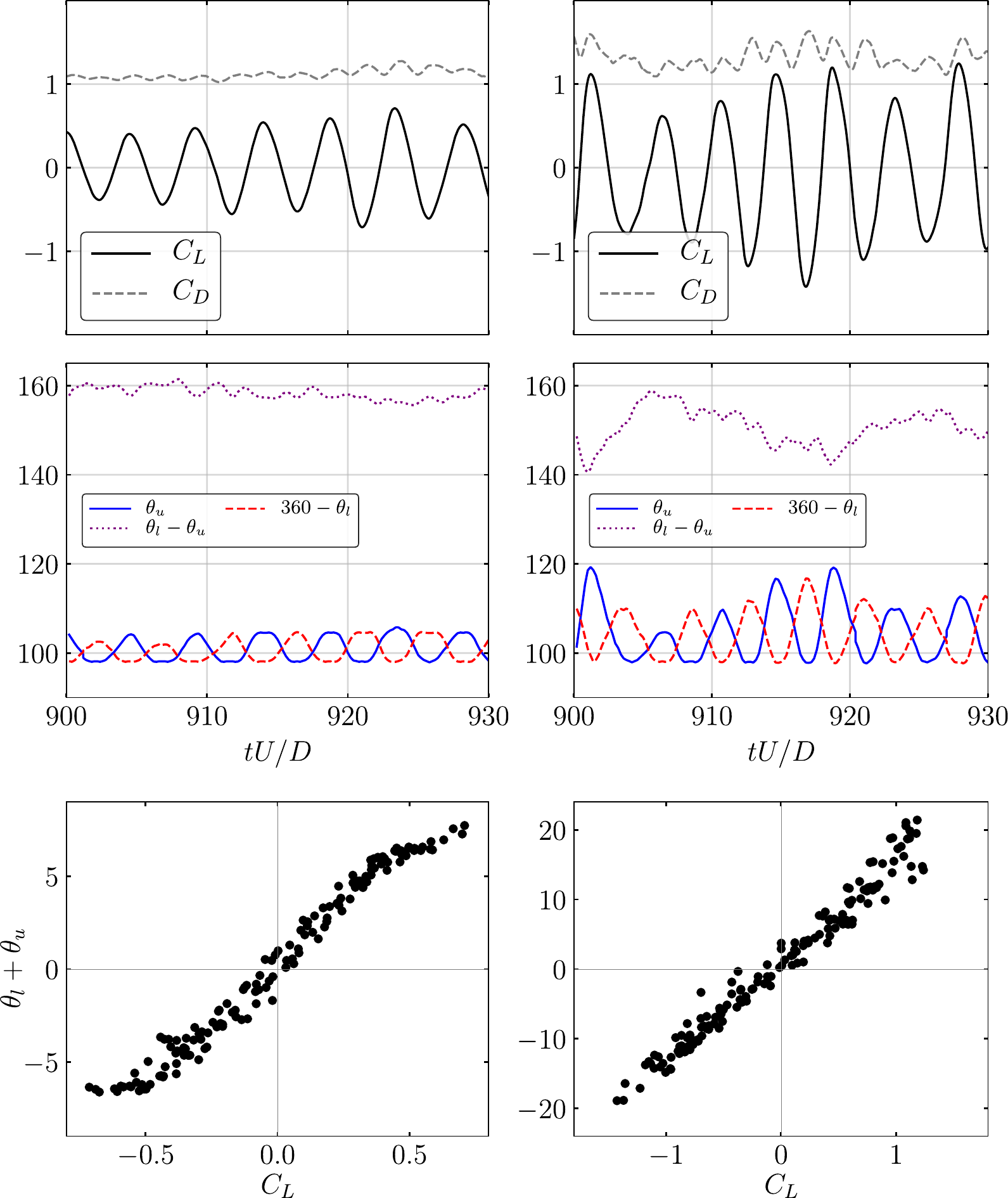}}
  \caption{Top: $C_L$ and $C_D$ temporal signals. The latter is calculated as $C_D=2F_x/(\rho U^2 D L_z)$ and the former as detailed in figure \ref{fig:TKE_CL} (without the r.m.s norm). Middle: Upper $\theta_u$ and lower $\theta_l$ separation angles temporal signal. The separation angle is calculated from the front stagnation point of the cylinder to the free shear layer separation point (vorticity practically 0 at the wall). Bottom: Correlation between the lift coefficient and the difference between $\theta_u$ and $360-\theta_l$. Left: $L_z=\pi$. Right: $L_z=0.5$.}
\label{fig:separation}
\end{figure}

On figure \ref{fig:separation}, the temporal signals of the lift and drag coefficients as well as the upper and lower separation angles of the free-shear layer are displayed for the $L_z=0.5$ and the $L_z=\pi$ cases. The oscillation of the separation points increases as the span is constricted inducing larger forces on the cylinder. A very high correlation of the upper and lower separation points angle with the lift coefficient is shown as well. Again, the lack of the Mode B instability on the constricted cases yields more coherent 2D vortical structures in the near wake region. Therefore, the small-scale 3D structures which normally dissipate most of the kinetic energy in 3D turbulence are not present leading to high-intensity vortices. This can be quantified by the enstrophy ($\Omega$) of the span-averaged spanwise vorticity in the near wake region $\mathcal{D}$ ($x \in [0.55, 2.1], y \in [-0.8, 0.8]$), which is time-averaged for the same temporal length as the body forces signals as follows,
\begin{equation}
\Omega = \int_\mathcal{D}\frac{1}{t_2-t_1}\int^{t_2}_{t_1}\frac{1}{L_z}\int^{L_z}_{0} \omega_z^{2} \,\,dz\,dt\,d\mathcal{D}.
\label{eq:enstrophy}
\end{equation}
It can be observed in table \ref{tab:enstrophy} that the enstrophy increases as the span is reduced, which can be understood as an increase of the rotational energy of the flow.

\begin{table}
  \begin{center}
\def~{\hphantom{0}}
  \begin{tabular}{lrrrrr}
       $L_z$                 & $\pi$  & $1$  & $0.5$ & $0.25$ & $0.1$ \\[3pt]
       $\Omega/\Omega_{\pi}$ & 1      & 1.32 & 1.71  & 2.07   & 2.47  \\
  \end{tabular}
  \caption{Near body enstrophy as defined in equation \ref{eq:enstrophy} for the different span cases. $\Omega_{\pi}$ is the enstrophy of the $L_z=\pi$ case.}
  \label{tab:enstrophy}
  \end{center}
\end{table}

The highly energised coherent vortices forming at the near wake region for the constricted cases induce a larger convective force on the free-shear layer. This translates to the large oscillations observed in figure \ref{fig:separation} and, ultimately, to the forces induced on the cylinder. With this, it can be argued that the coherent 2D structures have a greater impact on the forces induced to the cylinder than the 3D small-scale structures when both are present. 

\cite{Lumley1977} proposed the Lumley's triangle of turbulence which provides a way to classify the anisotropic state of turbulence. The anisotropy property of the Reynolds stress tensor can be extracted with,
\begin{equation}
b_{ij} = \frac{\overline{u_i' u_j'}}{\overline{u_k' u_k'}} - \frac{1}{3}\delta_{ij},
\end{equation}
where $\delta$ is the Kronecker delta and $b_{ij}$ is the anisotropic Reynolds stress tensor (which evidently vanishes for isotropic turbulence). This dimensionless and traceless tensor has two non-zero invariants, $II=-b_{ij}b_{ji}/2$ and $III=b_{ij}b_{jk}b_{ki}/3$. These invariants are often rewritten as $\eta^2=-II/3$ and $\xi^3=III/2$ to better appreciate the nonlinear behaviour of the trajectories of return to isotropy of homogeneous turbulence \citep{Choi2001}. It has been shown that all the possible turbulence states are mapped within the triangle \citep{Lumley1977, Lumley1978}.

The different states of turbulence are classified in the triangle as follows: The top right elbow indicates a one-dimensional (or one component) state with a single non-zero eigenvalue (the eigenvalues can be understood in physical terms as the normal stresses in the principal axes of the anisotropic Reynolds stress tensor). The top left elbow indicates a 2D isotropic state where one eigenvalue vanishes and the two remaining are equal. The top curve connecting the elbows indicates a 2D turbulence state where one eigenvalue vanishes and the addition of the two remaining eigenvalues is constant. The left and right straight lines correspond to a negative or positive $\left( \xi \right)$ axisymmetric state since one eigenvalue is smaller than the other two (which are equal) or greater than the other two (which are equal) respectively. Finally, the $(0,0)$ point indicates 3D isotropic turbulence since all of the anisotropic tensor eigenvalues vanish.

\begin{figure}
	\vspace*{0.3cm}
  \centerline{\includegraphics[width=1\textwidth]{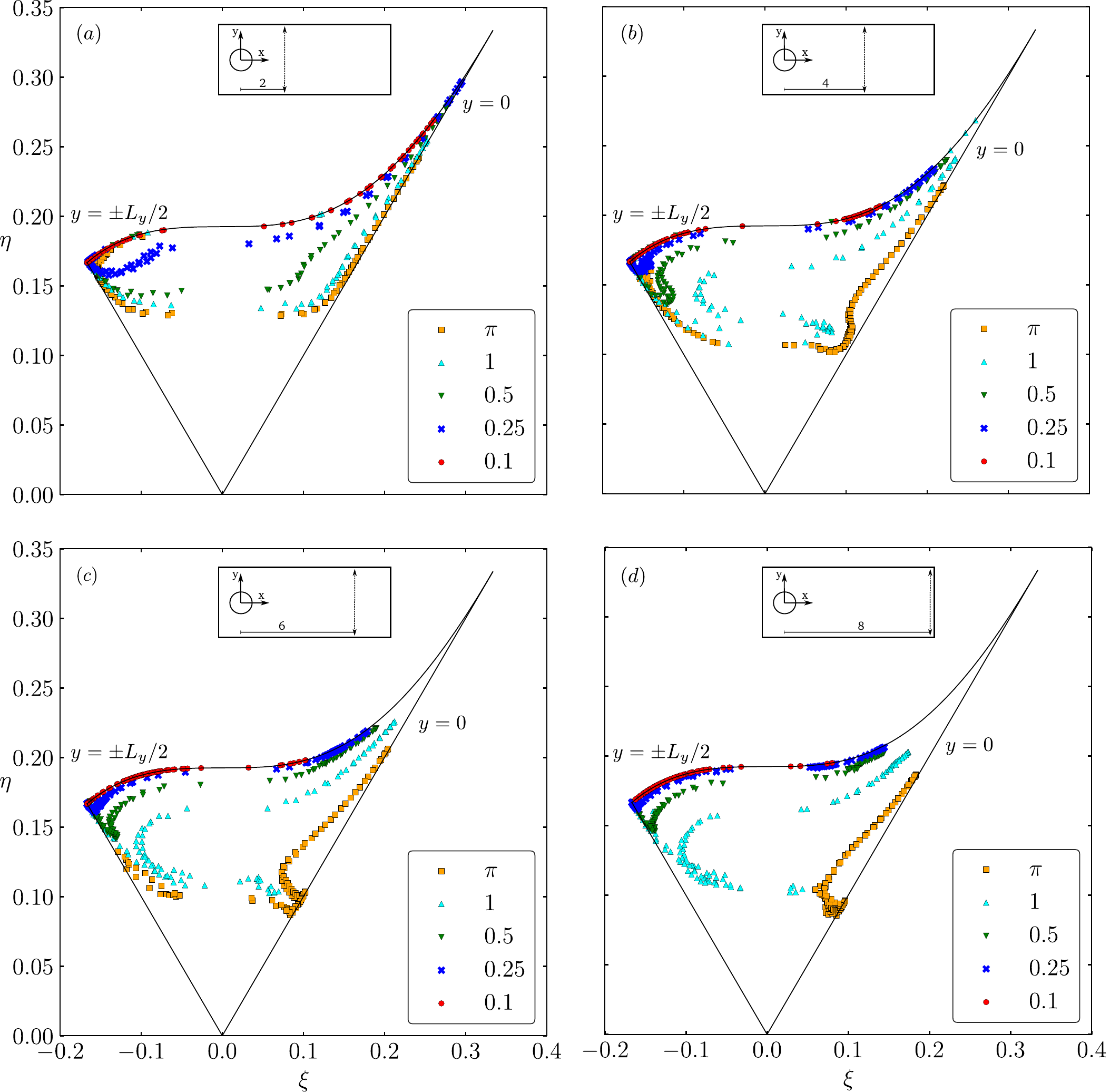}}
  \caption{Lumley's triangle constructed across the wake width at: $(a)$ $x=2$, $(b)$ $x=4$, $(c)$ $x=6$, $(d)$ $x=8$.}
\label{fig:lumleys_triangle}
\end{figure}

Figure \ref{fig:lumleys_triangle} displays the Lumley's triangle constructed across the wake width at different $x$ locations. Hence, every point in the triangle corresponds to a point in the domain and the collection of points for each case represents the collection of vertically aligned points in the domain at different $x$ locations on the wake. It can be appreciated that all of the points are located within the triangle, therefore all of computed Reynolds stresses are realisable (\textit{i.e.} have positive and real eigenvalues). In general, most of the cases transition from one state to another state of turbulence across the wake width. Only for the almost 2D case $\left(L_z=0.1\right)$, the state of turbulence is always 2D (or two-component), since $\overline{w'w'}$ is effectively negligible everywhere.

As the triangle is constructed further downstream on the wake, the trajectories (or collection of points) of the $L_z=0.25$ and $L_z=0.5$ cases move closer to the 2D turbulence state location (upper curve). This emphasises once again the wake two-dimensionalisation caused by the large-scale vortical structures on cases with critical span which contain a cross-over of 2D and 3D turbulence as shown in previous results. In contrast, the $L_z=1$ and $L_z=\pi$ cases remain approximately at the same $\eta$ region showing that the two-dimensionalisation is not as effective.

Additionally, the trajectories present a negative axisymmetric almost two-component state for the locations far from the wake centreline, \textit{i.e.} the location close to $y=\pm L_y/2$, since one of the normal turbulent stresses $\overline{w'w'}$ is smaller than the other ones. On the centreline, $\overline{v'v'}$ is larger than the other stresses causing a shift to the positive axisymmetric state.

A comparison of the different cases at the same $x$ location also shows a more noticeable 2D turbulence state as the span is constricted. This difference is less noticeable close to the cylinder (figure \ref{fig:lumleys_triangle}a), where even the $L_z=0.5$ case trajectory resembles the $L_z =\pi$ case. Again, this evidences that 3D turbulence is present close to the wall even in considerably constricted cases.

Consider now the vorticity transport equation (VTE) which can be written as,
\begin{equation}
\partial_{t}\boldsymbol\omega+\mathbf{u}\cdot\nabla \boldsymbol\omega=\boldsymbol\omega\cdot\nabla\mathbf{u}+\Rey^{-1}\nabla^2\boldsymbol\omega,
\end{equation}
where $\boldsymbol\omega\left(\mathbf{x}, t\right) = \left(\omega_x, \omega_y, \omega_z\right)$ is the vorticity vector field defined as $\boldsymbol\omega=\nabla\times\mathbf{u}$. The vortex-stretching term, $\boldsymbol\omega\cdot\nabla\mathbf{u}$, is often pointed as the term responsible for the direct energy cascade of the TKE. The stretching of a vortex tube causes a reduction on its diameter while increasing the rotation speed of the vortex by conservation of angular momentum. This term vanishes in the 2D formulation of the VTE since the stretch of the vortex tube is perpendicular to the plane of rotation. From this mathematical and physical difference, different turbulence dynamics are captured on 2D or 3D computations and, therefore, 3D turbulence can be directly linked to this term. 

\begin{figure}
	\vspace*{0.3cm}
  \centerline{\includegraphics[width=1\textwidth]{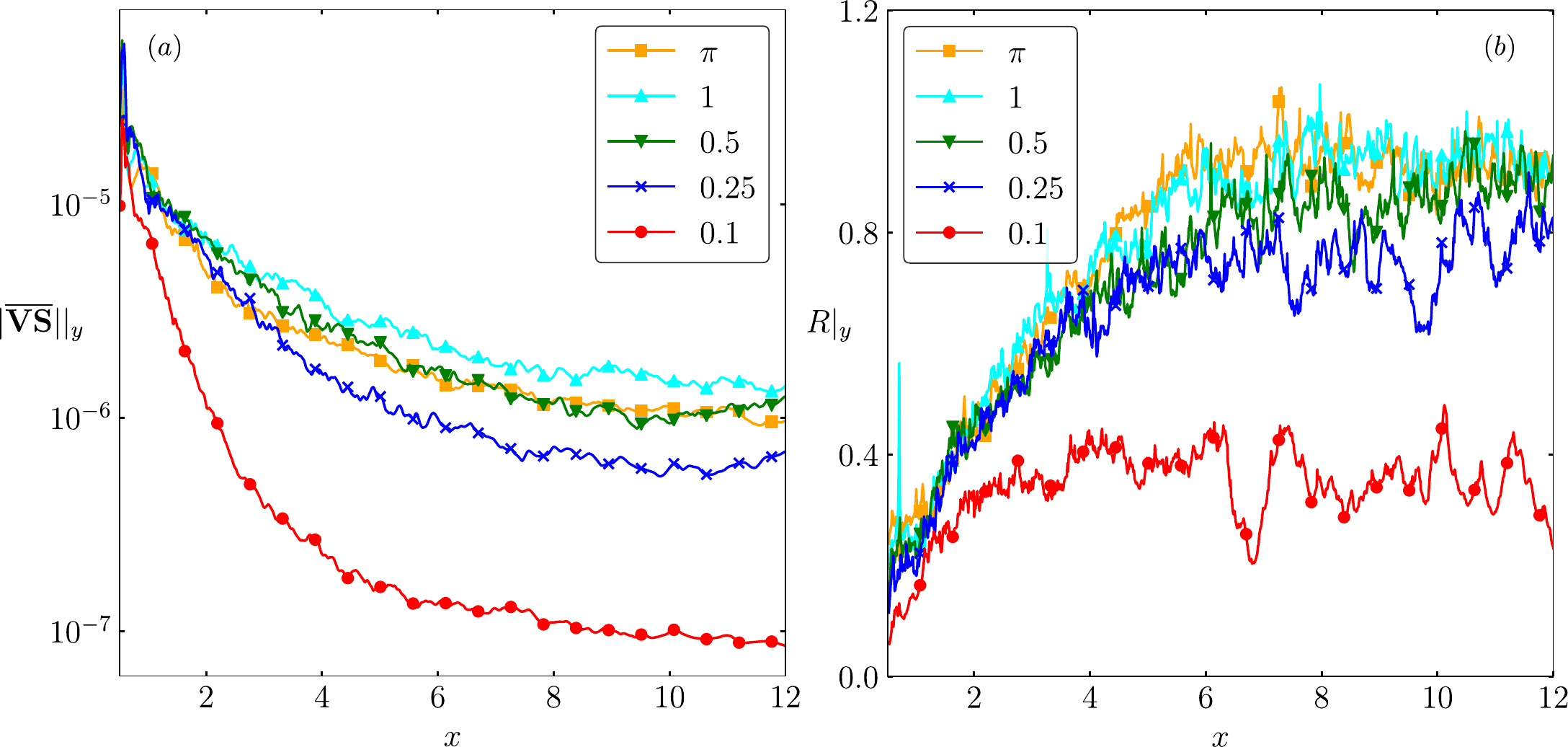}}
  \caption{$(a)$ Modulus of the mean vortex-stretching term averaged along the vertical direction. $(b)$ Ratio between the modulus of the mean vortex-stretching term and the modulus of the mean vortex-advecting term averaged along the vertical direction.}
\label{fig:vortex_stretching}
\end{figure}

Figure \ref{fig:vortex_stretching} displays the modulus of the mean vortex-stretching term and its ratio $R$ with the mean vortex-advecting term. These quantities are averaged on the vertical direction. On figure \ref{fig:vortex_stretching}a, it can be observed that the vortex-stretching term decays faster while moving downstream from the cylinder for cases with shorter span. The vortex-stretching decay demonstrates again the two-dimensionalisation of the flow by the large-scale vortices. The ratio $R$ on figure \ref{fig:vortex_stretching}b shows that the vortex-advecting term decays faster (because of the wake momentum deficit) than the vortex-stretching term along the streamwise direction up to $x=6$ where the ratio is kept constant. It can also be observed that the vortex-stretching term becomes as important as the vortex-advecting term with increasing span.

\section{Conclusion}\label{Conclusion}
The span effect on the turbulence dynamics of a flow past a circular cylinder at $\Rey=10000$ has been investigated using spectras  and two-point correlation on different locations in the domain, the TKE along the wake, the separation points, the Lumley's triangle of turbulence and the mean vortex-stretching term of the VTE. It has been shown that 3D turbulence is present even for highly constricted cases (for example $L_z=0.25$) which is generated by the cylinder wall (figure \ref{fig:velocity_spectras}a). However, the small-scale structures rapidly get two-dimensionalised by the large-scale K\'{a}rm\'{a}n vortices when the span is 50\% of the diameter or less. This is linked to the Mode B instability wavelength being longer than the periodic span, in agreement with \cite{Bao2016}. Since the Mode B instability helps sustaining the turbulent structures advected from the shear layer, the lack of it prevents large-scale 3D structures to be created and less dissipative structures can be sustained. In this scenario, 2D turbulence takes over and dominates the wake dynamics creating larger, stronger and more coherent vortices. Ultimately, the coherent and energised vortical structures induce a larger convective force on the free-shear layer. This translates to larger oscillations and, finally, higher forces on the cylinder.

The flow turbulence transition from 3D to 2D caused by a geometry constriction found in this work is in agreement with the physical mechanisms described in the obstacle-free turbulence work of \cite{Smith1996}, \cite{Celani2010} and \cite{Biancofiore2014}. In the present study with solid boundaries, the main difference is found on the presence of small-scale 3D turbulence even in highly-constricted geometries $\left(L_z=0.25\right)$ which leads to a coexistence of 2D and 3D turbulence close to the cylinder wall. This is observed not only in the $L_z=0.25$ case but also for the $L_z=0.5$ and $L_z=1$ cases as shown in figures \ref{fig:velocity_spectras}b and \ref{fig:velocity_spectras}c respectively. Note that the crossover between 2D and 3D turbulence dynamics arises in different points in the spatial domain depending on the span length. The shorter the span, the closer to the cylinder it takes place evidencing that the wake two-dimensionalisation transitions at different locations in function of the domain geometric anisotropy.

On the other hand, a very rapid two-dimensionalisation is found in the present cases because of the natural large-scale rotation motion of the K\'{a}rm\'{a}n vortices. A large-scale rotation as a mechanism of two-dimensionalisation has been also found in other works such as \cite{Smith1996} and \cite{Xia2011}. These two mechanisms combined yield to a rapid transition from the 3D to 2D turbulence dynamics when the span is shorter than the Mode B instability wavelength.

\section*{Acknowledgements}
The authors acknowledge the use of the IRIDIS High Performance Computing Facility, and associated support services at the University of Southampton, in the completion of this work. The authors also acknowledge the support of the Singapore Agency for Science, Technology and Research (A*STAR) Research Attachment Programme (ARAP) as a collaboration between the A*STAR Institute of High Performance Computing (IHPC) and the Faculty of Engineering and the Environment of the University of Southampton.

\appendix
\section{}\label{appA}
A grid convergence study has been conducted in order verify the correct implementation of the governing equations and to show that the selected grid resolution and averaging time length is suitable for a proper analysis. Three grids with different resolution $(\Delta)$ designed as shown in figure \ref{fig:computational_domain} have been investigated for the $L_z=\pi$ case (see details in table \ref{tab:verification_grids}). The refinement ratio is $\sqrt{2}$.

\begin{figure}
	\vspace*{0.3cm}
  \centerline{\includegraphics[width=0.94\textwidth]{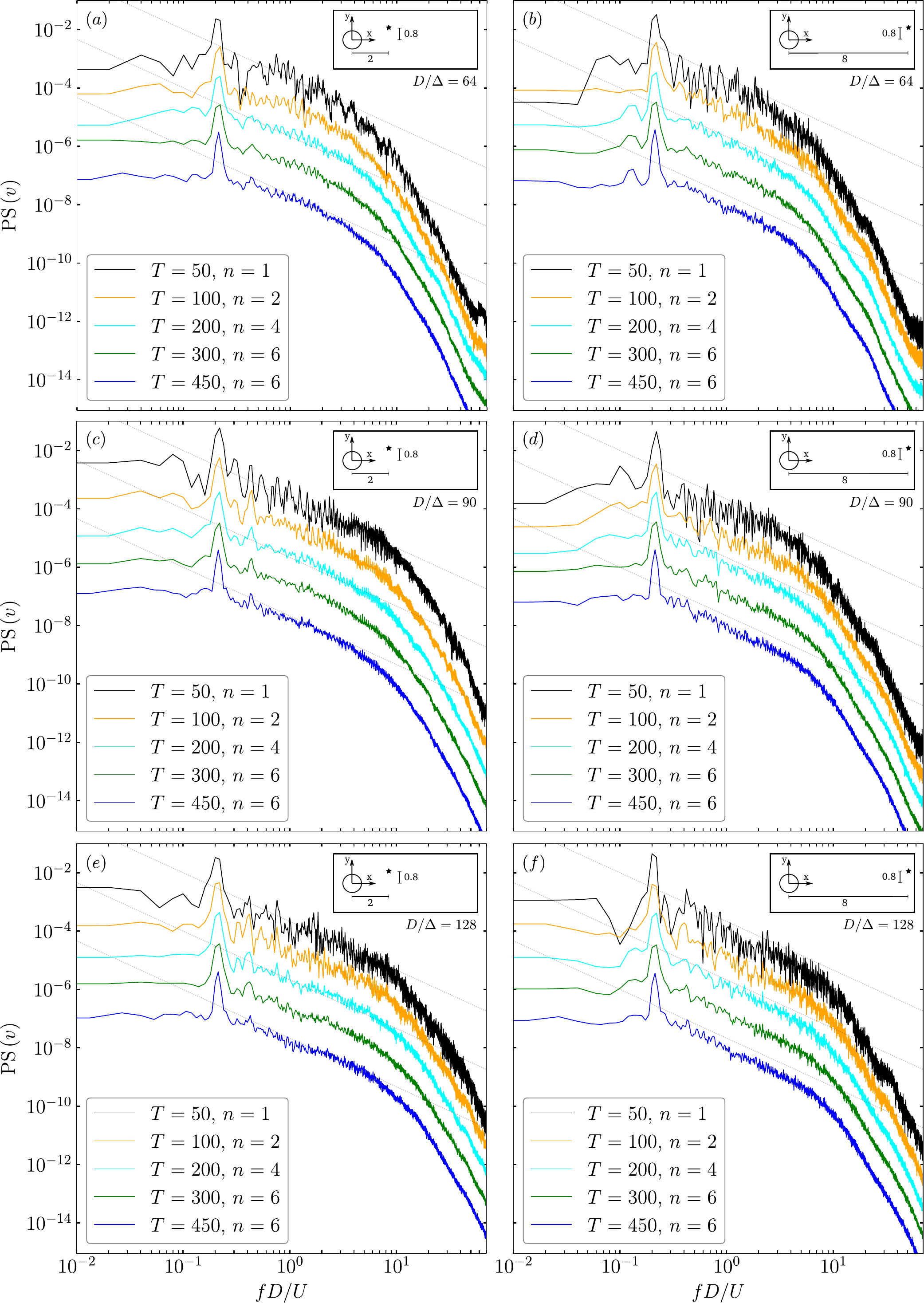}}
  \caption{Vertical velocity component temporal power spectra at the close (left) and far (right) wake regions. The coarse, medium and fine grid results are displayed at the first, second and third row respectively. The spectras are shifted a factor of 10 and the vertical axis ticks correspond to the $T=50$ case. The number of the time signal splits selected for the Welch method ($n$) ensures a minimum of $50T$ per split (at least 8 times larger than the lowest frequency of interest, \textit{i.e.} the shedding frequency which is around $5T$) and a maximum of 6 splits per total time signal. The dotted line corresponds to a $-5/3$ slope.}
\label{fig:velocity_spectras_GC}
\end{figure}

\begin{figure}
  \vspace*{0.0cm}
  \centerline{\includegraphics[width=1\textwidth]{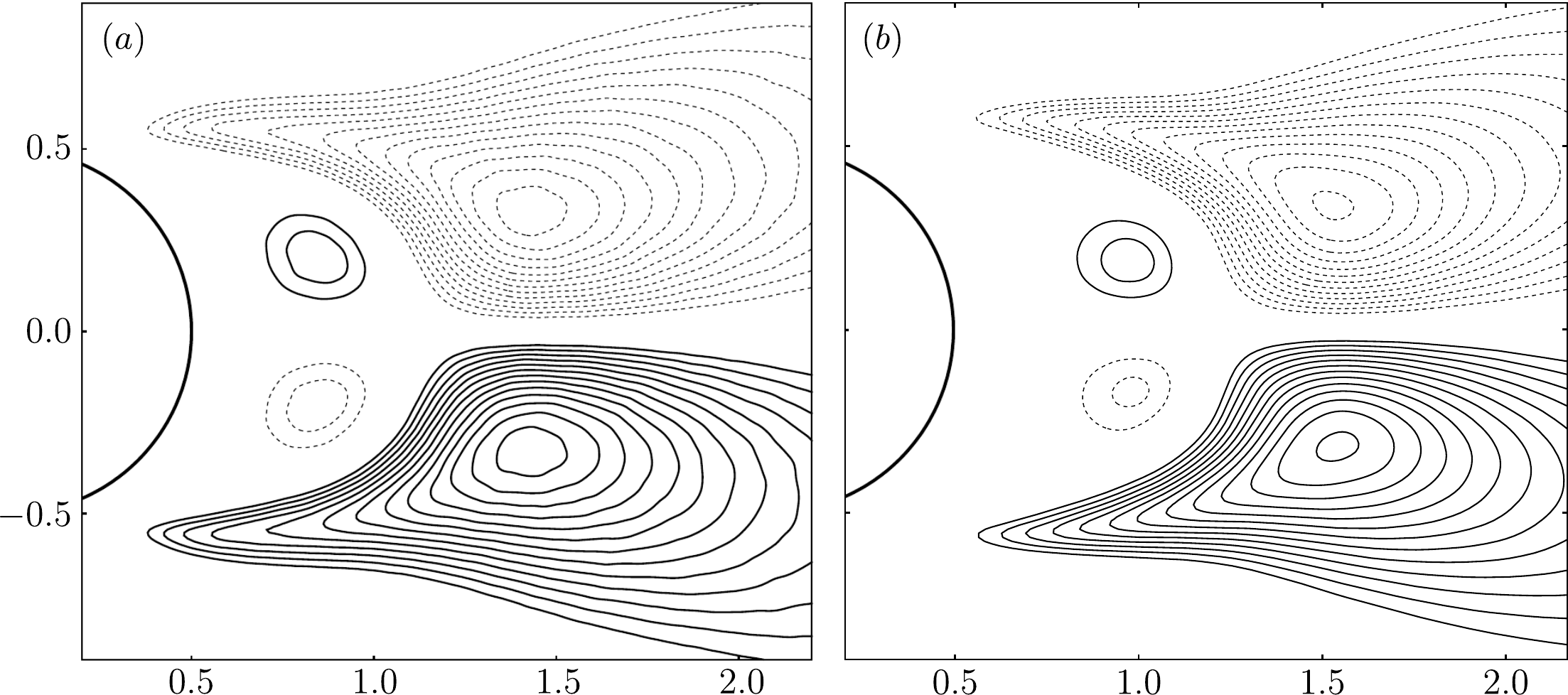}}
  \caption{Comparison of the Reynolds shear stress $\overline{u'v'}$ between: (a) DNS data from \cite{Dong2005} and (b) present solver. The contour levels criteria is the same as the reference data: ${|\overline{u'v'}|}_{\mathrm{min}}=0.03$ and $\Delta\overline{u'v'}=0.01$. Positive and negative levels are noted with continuous and dashed lines respectively.}
\label{fig:uv_validation}
\end{figure}

The quantities of interest of the present work are the turbulence statistics arising from time and spatial averages. Because of this, we need to ensure that the presented results are statistically stationary and have converged in spatial resolution terms. The velocity temporal power spectra constructed analogously to figure \ref{fig:velocity_spectras} at the $\left(2,0.8\right)$ and $\left(8,0.8\right)$ locations is used with this purpose. The spectra produced by investigated grids is shown for different time signal lengths in figure \ref{fig:velocity_spectras_GC}.

\begin{table}
  \begin{center}
\def~{\hphantom{0}}
  \begin{tabular}{rrr}
       $D/\Delta$ & $N_{CG}$ (M) & $N_{T}$ (M) \\[3pt]
       64  & 65.3 & 109.2\\
       90  & 189.5 & 311.4 \\
       128 & 510.6 & 855.6\\
  \end{tabular}
  \caption{Details of the three different grids ranging from coarse ($D/\Delta=64$, \textit{i.e.} 64 cells per diameter) to fine ($D/\Delta=128)$ for $L_z=\pi$. $N_{GC}$ refers to the total number of cells (in millions) for the Cartesian grid subdomain. $N_T$ refers to the total number of cells (in millions).}
  \label{tab:verification_grids}
  \end{center}
\end{table}

At the near wake region, the coarse grid shows a converged spectra ($-5/3$ inertial subrange decay) for time signals longer than 200T. A similar behaviour is captured with the medium grid, which shows a converged spectra for time signals longer than 100T and the resulting inertial subrange accommodates more frequencies. At the far wake region, the coarse grid presents slightly steeper spectras than at the close wake region. This can be caused by an insufficient resolution in span, which would induce a two-dimensionalisation effect. On the other hand, the medium grid still presents converged spectras for time signals longer than 100T. The fine grid leads to results very similar to the medium grid. Hence, it is shown that using the medium grid with simulation times over 100T (500T is used in the results presented in this work) provides statistically stationary results both in the close and far wake regions.

To validate the wake turbulence statistics, the Reynolds shear stress $\overline{u'v'}$ has also been qualitatively analysed. It can be observed in figure \ref{fig:uv_validation} that the shear stress predicted by the present solver is in very good agreement with the Direct Numerical Simulation (DNS) data from \cite{Dong2005}, with just a slight shift of the field structures on the streamwise direction. All positive and negative regions are correctly captured which display an antisymmetric pattern with respect to the centreline of the wake.

\bibliographystyle{jfm}
\bibliography{main}

\begin{thebibliography}{33}
\expandafter\ifx\csname natexlab\endcsname\relax\def\natexlab#1{#1}\fi
\def\au#1{#1} \def\ed#1{#1} \def\yr#1{#1}\def\at#1{#1}\def\jt#1{\textit{#1}}
  \def\bt#1{#1}\def\bvol#1{\textbf{#1}} \def\vol#1{#1} \def\pg#1{#1}
  \def\publ#1{#1}\def\arxiv#1{#1}\def\org#1{#1}\def\st#1{\textit{#1}}

\bibitem[Adams \& Hickel(2009)]{Adams2009}
{\sc \au{Adams, N.~A.} \& \au{Hickel, S.}} \yr{2009} Implicit large-eddy
  simulation: Theory and application.  \bt{In {\em Advances in Turbulence
  XII\/}},  \pg{pp. 743--750}.  \publ{Berlin, Heidelberg: Springer Berlin
  Heidelberg}.

\bibitem[Bao {\em et~al.\/}(2016)Bao, Palacios, Graham \& Sherwin]{Bao2016}
{\sc \au{Bao, Y.}, \au{Palacios, R.}, \au{Graham, J. M.~R.} \& \au{Sherwin,
  S.}} \yr{2016}  \at{Generalized thick strip modelling for vortex-induced
  vibration of long flexible cylinders}.  \jt{J.~Comp. Phys.}  \bvol{321},
  \pg{1079--1097}.

\bibitem[Bao {\em et~al.\/}(2019)Bao, Zhu, Huan, Wang, Zhou, Han, Palacios,
  Graham \& Sherwin]{Bao2019}
{\sc \au{Bao, Y.}, \au{Zhu, H.B.}, \au{Huan, P.}, \au{Wang, R.}, \au{Zhou, D.},
  \au{Han, Z.L.}, \au{Palacios, R.}, \au{Graham, M.} \& \au{Sherwin, S.}}
  \yr{2019}  \at{Numerical prediction of vortex-induced vibration of flexible
  riser with thick strip method}.  \jt{J.~ Fluids and Struct.} .

\bibitem[Batchelor(1969)]{Batchelor1969}
{\sc \au{Batchelor, G.~K.}} \yr{1969}  \at{Computation of the energy spectrum
  in homogeneous two-dimensional turbulence}.  \jt{Phys. Fluids}
  \bvol{12}~(12),  \pg{II--233--II--239}.

\bibitem[Biancofiore(2014)]{Biancofiore2014}
{\sc \au{Biancofiore, L.}} \yr{2014}  \at{Crossover between two- and
  three-dimensional turbulence in spatial mixing layers}.  \jt{J.~Fluid Mech.}
  \bvol{745},  \pg{164--179}.

\bibitem[Biancofiore {\em et~al.\/}(2012)Biancofiore, Gallaire \&
  Pasquetti]{Biancofiore2012}
{\sc \au{Biancofiore, L.}, \au{Gallaire, F.} \& \au{Pasquetti, R.}} \yr{2012}
  \at{Influence of confinement on obstacle-free turbulent wakes}.
  \jt{Computers \& Fluids}  \bvol{58},  \pg{27--44}.

\bibitem[Bloor(1964)]{Bloor1964}
{\sc \au{Bloor, M.~Susan}} \yr{1964}  \at{The transition to turbulence in the
  wake of a circular cylinder}.  \jt{J.~Fluid Mech.}  \bvol{19}~(2),
  \pg{290--304}.

\bibitem[Boffetta \& Ecke(2012)]{Boffetta2012}
{\sc \au{Boffetta, G.} \& \au{Ecke, R.~E.}} \yr{2012}  \at{Two-dimensional
  turbulence}.  \jt{Annu. Rev. Fluid Mech.}  \bvol{44},  \pg{427--451}.

\bibitem[Celani {\em et~al.\/}(2010)Celani, Musacchio \& Vincenzi]{Celani2010}
{\sc \au{Celani, A.}, \au{Musacchio, S.} \& \au{Vincenzi, D.}} \yr{2010}
  \at{Turbulence in more than two and less than three dimensions}.  \jt{Phys.
  Rev. Lett.}  \bvol{104}~(18),  \pg{184506}.

\bibitem[Choi \& Lumley(2001)]{Choi2001}
{\sc \au{Choi, K.-S.} \& \au{Lumley, J.~L.}} \yr{2001}  \at{The return to
  isotropy of homogeneous turbulence}.  \jt{J.~Fluid Mech.}  \bvol{436},
  \pg{59--84}.

\bibitem[Chyu \& Rockwell(1996)]{Chyu1996}
{\sc \au{Chyu, C.} \& \au{Rockwell, D.}} \yr{1996}  \at{Evolution of patterns
  of streamwise vorticity in the turbulent near wake of a circular cylinder}.
  \jt{J.~Fluid Mech.}  \bvol{320},  \pg{117--137}.

\bibitem[Dong \& Karniadakis(2005)]{Dong2005}
{\sc \au{Dong, S.} \& \au{Karniadakis, G.~E.}} \yr{2005}  \at{{DNS} of flow
  past a stationary and oscillating cylinder at {R}e=10000}.  \jt{J.~Fluids and
  Struct.}  \bvol{20},  \pg{519--531}.

\bibitem[Dritschel {\em et~al.\/}(2008)Dritschel, Scott, Macaskill, Gottwald \&
  Tran]{Dritschel2008}
{\sc \au{Dritschel, D.~G.}, \au{Scott, R.~K.}, \au{Macaskill, C.},
  \au{Gottwald, G.~A.} \& \au{Tran, C.~V.}} \yr{2008}  \at{Unifying scaling
  theory for vortex dynamics in two-dimensional turbulence}.  \jt{Phys. Rev.
  Lett.}  \bvol{101},  \pg{094501}.

\bibitem[Gilbert(1988)]{Gilbert1988}
{\sc \au{Gilbert, A.~D.}} \yr{1988}  \at{Spiral structures and spectra in
  two-dimensional turbulence}.  \jt{J.~Fluid Mech.}  \bvol{193},
  \pg{475--497}.

\bibitem[Hendrickson {\em et~al.\/}(2019)Hendrickson, Weymouth \&
  Yue]{Hendrickson2019}
{\sc \au{Hendrickson, K.}, \au{Weymouth, G.~D.} \& \au{Yue, D. K.-P. Yue~Yu,
  X.}} \yr{2019}  \at{Wake behind a three-dimensional dry transom stern. part
  1: Flow structure and large-scale air entrainment}.  \jt{J.~Fluid Mech} .

\bibitem[Kourta {\em et~al.\/}(1987)Kourta, Boisson, Chassaing \& {Ha
  Minh}]{Kourta1987}
{\sc \au{Kourta, A.}, \au{Boisson, H.~C.}, \au{Chassaing, P.} \& \au{{Ha Minh},
  H.}} \yr{1987}  \at{Nonlinear interaction and the transition to turbulence in
  the wake of a circular cylinder}.  \jt{J.~Fluid Mech.}  \bvol{181},
  \pg{141--161}.

\bibitem[Kraichnan(1967)]{Kraichnan1967}
{\sc \au{Kraichnan, R.~H.}} \yr{1967}  \at{Inertial ranges in two-dimensional
  turbulence}.  \jt{Phys. Fluids}  \bvol{10},  \pg{1417--1423}.

\bibitem[Leith(1968)]{Leith1968}
{\sc \au{Leith, C.~E.}} \yr{1968}  \at{Diffusion approximation for
  two-dimensional turbulence}.  \jt{Phys. Fluids}  \bvol{11},  \pg{671--673}.

\bibitem[Lumley(1978)]{Lumley1978}
{\sc \au{Lumley, J.~L.}} \yr{1978}  \at{Computational modeling of turbulent
  flows}.  \jt{Adv. Appl. Mech.}  \bvol{18},  \pg{123--176}.

\bibitem[Lumley \& Newman(1977)]{Lumley1977}
{\sc \au{Lumley, J.~L.} \& \au{Newman, G.~R.}} \yr{1977}  \at{The return to
  isotropy of homogeneous turbulence}.  \jt{J.~Fluid Mech.}  \bvol{82},
  \pg{161--178}.

\bibitem[Maertens \& Weymouth(2015)]{Maertens2015}
{\sc \au{Maertens, A.~P.} \& \au{Weymouth, G.~D.}} \yr{2015}  \at{Accurate
  {C}artesian-grid simulations of near-body flows at intermediate {R}eynolds
  numbers}.  \jt{Comput. Methods Appl. Mech. Engrg.}  \bvol{283},
  \pg{106--129}.

\bibitem[Mittal \& Balachandar(1995)]{Mittal1995}
{\sc \au{Mittal, R.} \& \au{Balachandar, S.}} \yr{1995}  \at{Effect of
  three-dimensionality on the lift and drag of nominally two-dimensional
  cylinders}.  \jt{Phys. Fluids}  \bvol{7}~(8).

\bibitem[Noack(1999)]{Noack1999}
{\sc \au{Noack, B.~R.}} \yr{1999}  \at{On the flow around a circular cylinder.
  {P}art {I}: laminar and transitional regime}.  \jt{J.~Appl. Math. and Mech.}
  \bvol{79},  \pg{223--226}.

\bibitem[Norberg(2003)]{Norberg2003}
{\sc \au{Norberg, C.}} \yr{2003}  \at{Fluctuating lift on a circular cylinder:
  review and new measurements}.  \jt{J.~Fluids and Struct.}  \bvol{17},
  \pg{57--96}.

\bibitem[Pope(2000)]{Pope2000}
{\sc \au{Pope, S.~B.}} \yr{2000} {\em Turbulent Flows\/}.  \publ{Cambridge
  University Press}.

\bibitem[Roshko(1954)]{Roshko1954}
{\sc \au{Roshko, A.}} \yr{1954}  \bt{On the development of turbulent wakes from
  vortex streets}. NACA Report 1191.  \org{National Advisory Committee for
  Aeronautics}, Washington D.C.

\bibitem[Schulmeister {\em et~al.\/}(2017)Schulmeister, Dahl, Weymouth \&
  Triantafyllou]{Schulmeister2017}
{\sc \au{Schulmeister, James~C.}, \au{Dahl, J.~M.}, \au{Weymouth, G.~D.} \&
  \au{Triantafyllou, M.~S.}} \yr{2017}  \at{Flow control with rotating
  cylinders}.  \jt{J.~Fluid Mech.}  \bvol{825},  \pg{743--763}.

\bibitem[Smith {\em et~al.\/}(1996)Smith, Chasnov \& Waleffe]{Smith1996}
{\sc \au{Smith, L.~M.}, \au{Chasnov, J.~R.} \& \au{Waleffe, F.}} \yr{1996}
  \at{Crossover from two- to three-dimensional turbulence}.  \jt{Phys. Rev.
  Lett.}  \bvol{77}~(12),  \pg{2467--2470}.

\bibitem[Weymouth \& Yue(2011)]{Weymouth2011}
{\sc \au{Weymouth, G.~D.} \& \au{Yue, D. K.~P.}} \yr{2011}  \at{Boundary data
  immersion method for {C}artesian-grid simulations of fluid-body interaction
  problems}.  \jt{J.~Comp. Phys.}  \bvol{230},  \pg{6233--6247}.

\bibitem[Williamson(1996{\natexlab{{\em a\/}}})]{Williamson1996b}
{\sc \au{Williamson, C. H.~K.}} \yr{1996{\natexlab{{\em a\/}}}}
  \at{Three-dimensional wake transition}.  \jt{J.~Fluid Mech.}  \bvol{328},
  \pg{345--407}.

\bibitem[Williamson(1996{\natexlab{{\em b\/}}})]{Williamson1996a}
{\sc \au{Williamson, C. H.~K.}} \yr{1996{\natexlab{{\em b\/}}}}  \at{Vortex
  dynamics in the cylinder wake}.  \jt{Annu. Rev. Fluid Mech.}  \bvol{28},
  \pg{477--539}.

\bibitem[{Xia} {\em et~al.\/}(2011){Xia}, {Byrne}, {Falkovich} \&
  {Shats}]{Xia2011}
{\sc \au{{Xia}, H.}, \au{{Byrne}, D.}, \au{{Falkovich}, G.} \& \au{{Shats},
  M.}} \yr{2011}  \at{Upscale energy transfer in thick turbulent fluid layers}.
   \jt{Nature Physics}  \bvol{7},  \pg{321--324}.

\bibitem[Xiao {\em et~al.\/}(2009)Xiao, Wan, Chen \& Eyink]{Xiao2009}
{\sc \au{Xiao, Z.}, \au{Wan, M.}, \au{Chen, S.} \& \au{Eyink, G.~L.}} \yr{2009}
   \at{Physical mechanism of the inverse energy cascade of two-dimensional
  turbulence: a numerical investigation}.  \jt{J.~Fluid Mech.}  \bvol{619},
  \pg{1--44}.

\end{thebibliography}
\end{document}